\documentclass{elsart5p}

\usepackage{natbib}

 \usepackage{graphicx}
\usepackage{psfig}
\usepackage{amssymb}

\begin{document}

\begin{frontmatter}



\title{A physical interpretation of the jet-like X-ray emission from supernova remnant W49B}


\author[label1,label2,label3]{M. Miceli}
\author[label1]{A. Decourchelle}
\author[label1]{J. Ballet}
\author[label3]{F. Bocchino}
\author[label4]{J. Hughes}
\author[label5,label6]{U. Hwang}
\author[label6]{R. Petre}

\address[label1]{DSM/DAPNIA/Service d'Astrophysique, AIM-UMR 7158, CEA Saclay, 91191 
Gif-sur-Yvette Cedex, France}
\address[label2]{Dipartimento di Scienze Fisiche ed Astronomiche, Sezione di 
Astronomia, Universit{\`a} di Palermo, Piazza del Parlamento 1, 90134 Palermo, 
Italy }
\address[label3]{INAF - Osservatorio Astronomico di Palermo, Piazza del Parlamento 1, 
90134 Palermo, Italy}
\address[label4]{Department of Physics and Astronomy, Rutgers University, 136 
Frelinghuysen Road, Piscataway, NJ 08854-8109, USA}
\address[label5]{Center for Astrophysical Sciences, The Johns Hopkins 
University, 3400 Charles St, Baltimore MD 21218}
\address[label6]{Laboratory for High Energy Astrophysics, Goddard Space Flight Center, 
Greenbelt, MD 20771, USA}
\begin{abstract}
In the framework of the study of supernova remnants and their complex interaction with the interstellar medium and the circumstellar material, we focus on the galactic supernova remnant W49B. Its morphology exhibits an X-ray bright elongated nebula, terminated on its eastern end by a sharp perpendicular structure aligned with the radio shell. The X-ray spectrum of W49B is characterized by strong K emission lines from Si, S, Ar, Ca and Fe. There is a variation of the temperature in the remnant with the highest temperature found in the eastern side and the lowest one in the western side. The analysis of the recent observations of W49B indicates that the remnant may be the result of an asymmetric bipolar explosion where the ejecta are collimated along a jet-like structure and the eastern jet is hotter and more Fe-rich than the western one. Another possible scenario associates the X-ray emission with a spherical explosion where parts of the ejecta are interacting with a dense belt of ambient material. To overcome this ambiguity we present new results of the analysis of an \emph{XMM-Newton} observation and we perform estimates of the mass and energy of the remnant. We conclude that the scenario of an anisotropic jet-like explosion explains quite naturally our observation results, but the association of W49B with a hypernova and a $\gamma-$ray burst, although still possible, is not directly supported by any evidence.
\end{abstract}


\end{frontmatter}

\section{Introduction}
\label{Introduction}

Supernova remnants (SNRs) are powerful sources of mass and energy for the interstellar medium (ISM) and their emission and evolution are deeply influenced by the complex interaction between the blast wave shock generated during the SN explosion and the ISM inhomogeneities. In the ejecta-dominated SNRs, the X-ray emission is associated with the ejected material expelled at supersonic speed during the SN explosion. The ejecta are heated up to X-ray emitting temperatures by the interaction with the reverse shock (\citealt{mck74}), or with the reflected shock, produced by the interaction of the remnant with a large ambient ISM cloud. The study of the young, ejecta-dominated SNRs is therefore a powerful diagnostic tool to study the ejecta, thus obtaining information on the dynamics of the supernova explosions and on their contribution to the chemical  evolution of the Galaxy. 

The galactic SNR W49B is one of the brightest ejecta-dominated SNRs observed in the X-ray band. It has a bright radio shell (whose diameter is $\sim 4'$) and a center filled morphology. Its distance $D_{W49B}$ is still uncertain. According to the HI absorption analysis of \citet{rgm72}, and considering the corrections by \citet{mr94}, $D_{W49B}\sim8$ kpc, but \citet{bt01} have shown that this estimate, although perfectly consistent with their new VLA data, is not supported by very clear evidence and that the range of possible $D_{W49B}$ values is quite large, being $\sim8$ kpc $<~D_{W49B}<$ $\sim12$ kpc. This upper limit is consistent with a possible association of W49B with the star forming region W49A (which is at $11.4$ kpc, according to \citealt{gmr92}), although there are a few indications that W49B is closer to us. In what follows, however, we will use $D=8$ kpc.

The X-ray emission of W49B has been widely studied and different observations have been performed with the past generations of X-ray satellites (\emph{Einstein Observatory}, \citealt{ptb84}; \emph{EXOSAT}, \citealt{spj85}, and $ASCA$, \citealt{fti95}, \citealt{hph00}). In particular, the analysis of the $ASCA$ spectrum of W49B has shown that the X-ray emission of W49B can be described with two thermal components of optically thin plasma with significant overabundances of Si, S, Ar, Ca, and Fe with respect to the
solar values (\citealt{hph00}).

In \citet{mdb06} (hereafter Paper I) we proposed two possible physical scenarios associated with W49B and both these interpretations are consistent with the observations subsequently analyzed by \citet{krr06}. The observed X-ray and IR emission of W49B can be the result of i) an asymmetric bipolar explosion (with the eastern jet being hotter and more Fe-rich than the western jet) or ii) a spherical explosion in a highly inhomogeneous circumstellar and interstellar medium.
Here we describe in detail these possible scenarios, we show new results of the analysis of the \emph{XMM-Newton} data presented in Paper I, and we try to obtain a more complete picture of the remnant and to discriminate between the two interpretations.

\section{The X-ray emitting plasma in W49B}
\label{X-rays}

\begin{figure}[htb!]
 \centerline{\hbox{     
     \psfig{figure=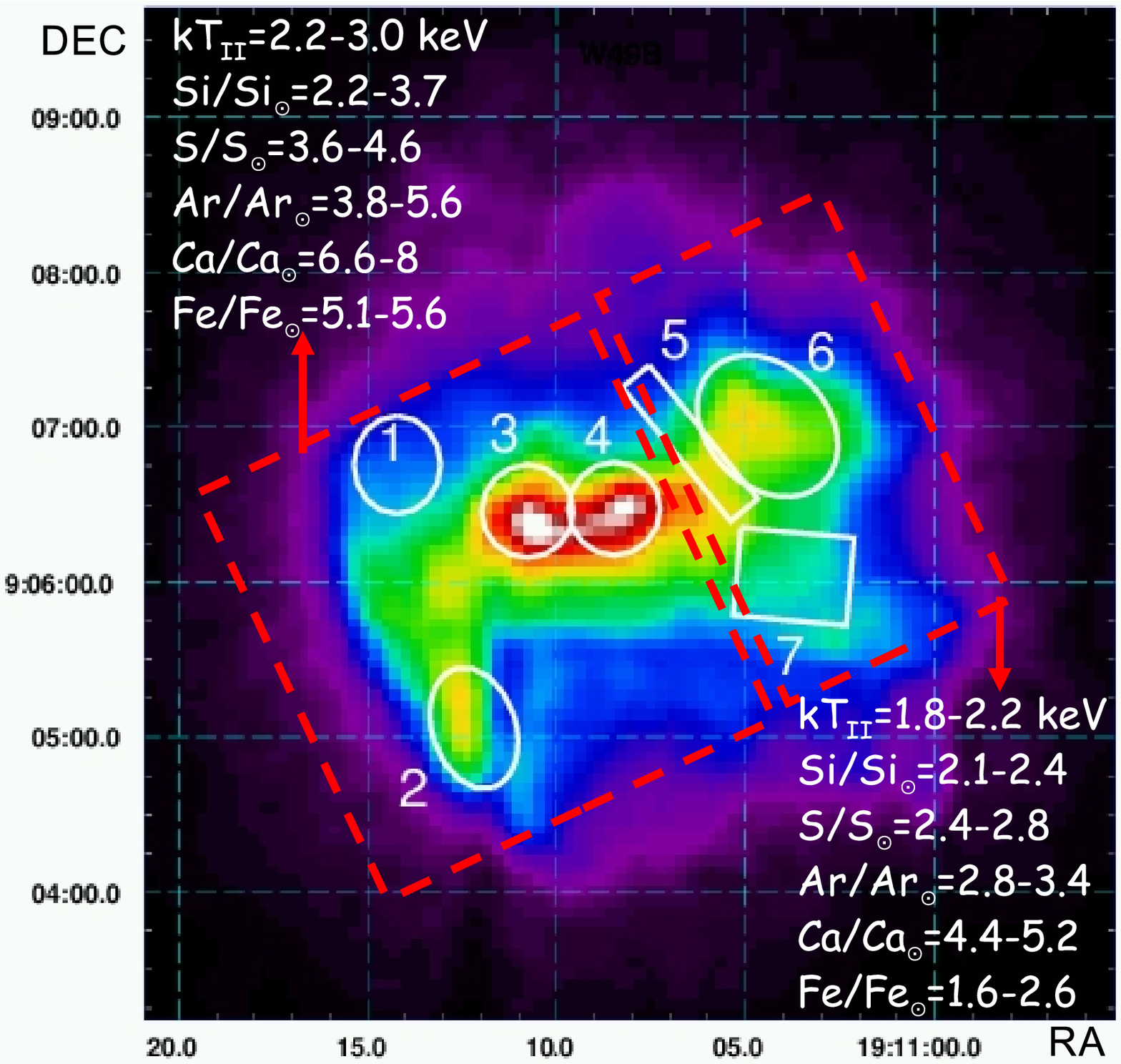,width=\columnwidth}
     }}
     \centerline{\hbox{     
     \psfig{figure=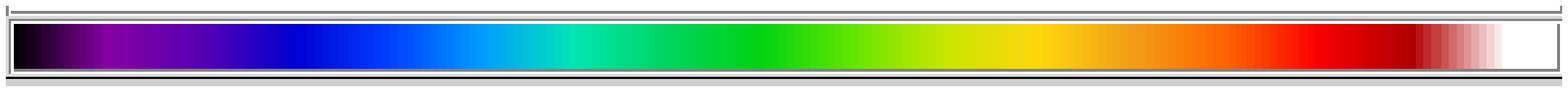,width=\columnwidth}
     }}     
\caption{\emph{XMM-Newton} EPIC count-rate image of W49B in the $1-9$ keV energy band. The image is vignetting-corrected, adaptively smoothed (with 
signal-to-noise ratio 10) and background-subtracted. The count-rate ranges between 0 and $7.2\times 10^{-3}$ s$^{-1}$. We have superimposed the ranges of the best fit parameter obtained in Paper I for the spectral regions at the central-eastern side of the remnant and for those at its western side. }
\label{fig:w49bspec}
\end{figure}

As shown in Fig. \ref{fig:w49bspec}, the supernova remnant W49B presents a complex X-ray morphology, which is characterized by a bright centrally elongated structure. This structure is terminated on the eastern side by a perpendicular sharp region and on the western side by a more diffuse, nearly aligned structure. The spatially resolved spectral analysis we performed in Paper I shows that the physical conditions of the plasma are not homogeneous throughout the remnant. All spectra are described well by two thermal components in collisional ionization equilibrium and lower temperatures are found in the western end of the remnant, while hotter regions are located in the center and at East. Everywhere in the remnant Si, S, Ar, Ca, and Fe are overabundant with respect to their solar values and we also measured a high Ni overabundance ($Ni/Ni_{\odot}=10^{+2}_{-1}$) in the central region of W49B.
While the central and eastern structure have similar abundances, the western region exhibits lower abundances. There is therefore an anisotropy in temperature and abundances, as shown in Fig. \ref{fig:w49bspec}, where we report the ranges of the best-fit parameters we derived in Paper I in the spectral regions at the central-eastern side of the remnant and in those at its western side. This anisotropy has been also confirmed by the analysis of the $Chandra$ observation described in \citet{krr06}. 

\begin{figure*}[htb!]
 \centerline{\hbox{     
     \psfig{figure=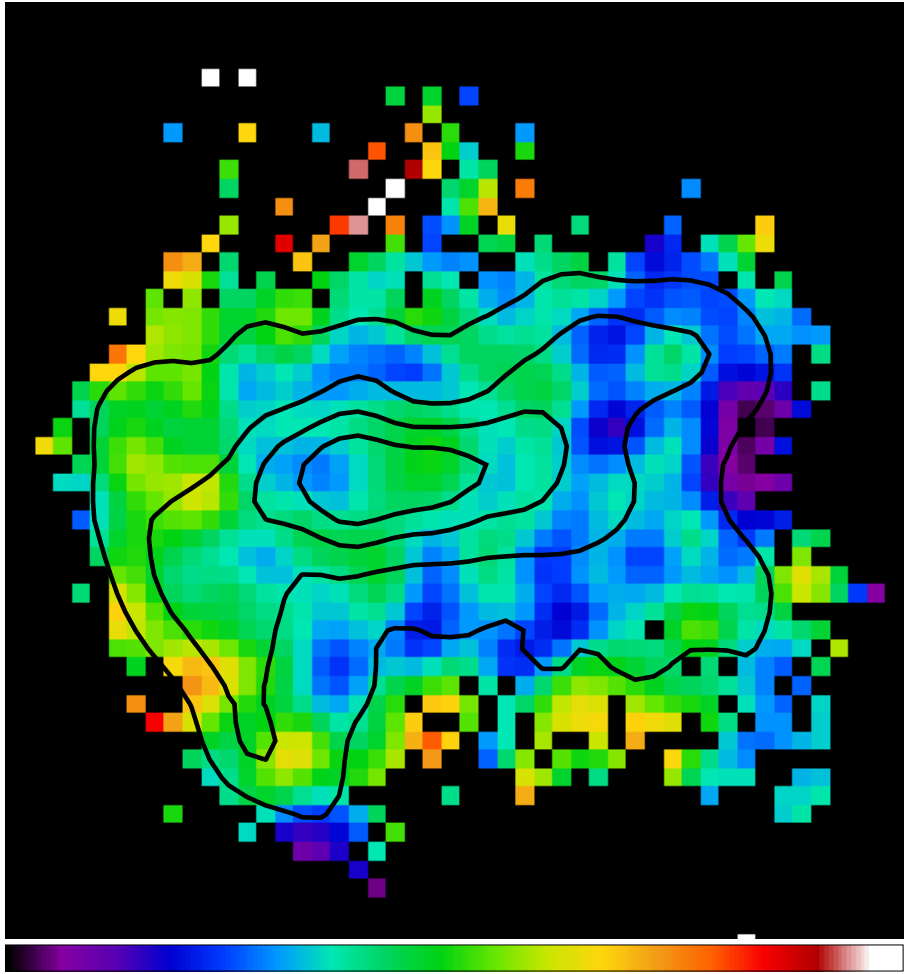,width=12 cm}
     }}
\caption{MOS mean photon energy map of the continuum emission (bin size$=6''$). In each pixel the local median photon energy of the photons detected by the MOS cameras in the continuum energy band $4.4-6.2$ keV is reported. We superimposed, in black, 4 contour levels indicating where the average is performed over more than 10, 20, 30, and 40 counts (from the outer to the inner level, respectively). Pixels with less than 4 counts have been masked out. The color bar has a linear scale and ranges between 4.85 keV and 5.25 keV.}
\label{fig:W49BMPE}
\end{figure*}

To highlight this temperature anisotropy and to obtain information about the thermal structure of the source we present here the MOS median photon energy (MPE) continuum map, that is an image where each pixel holds the median energy of the detected MOS photons in the continuum band $4.4-6.2$ keV. This map provides information about the spatial distribution of the spectral properties of the plasma, because the harder the spectrum (i. e. the higher the continuum temperature), the higher is the local MPE. To produce this map, we considered the \emph{XMM-Newton} EPIC MOS1 and MOS2 event files in the continuum energy band $4.4-6.2$ keV, with a bin size of $6''$ (so as to collect more than 10 counts per pixel everywhere in the remnant). For each pixel, we calculated the median energy of the photons and then we smoothed the map by using the following procedure: for each pixel $k$ of the MPE map we define its weight as $W_{k}=C_{k}~e^{\frac{-(i-k)^2}{\sigma^{2}}}$, where $C_{k}$ is the number of photons detected in that pixel, the $i$ index runs over all the other pixels and $\sigma=12''$. The smoothed median energy value at the $i$ pixel is then $SMPE_{i} = \Sigma_{k}~( W_{k}~MPE_{k} )/ \Sigma_{k}~ (MPE_{k})$. The smoothed  MPE map of W49B is shown in Fig. \ref{fig:W49BMPE} and confirms the presence of a higher continuum mean photon energy in the eastern side of the remnant than in its western end.

\begin{figure}[htb!]
 \centerline{\hbox{     
     \psfig{figure=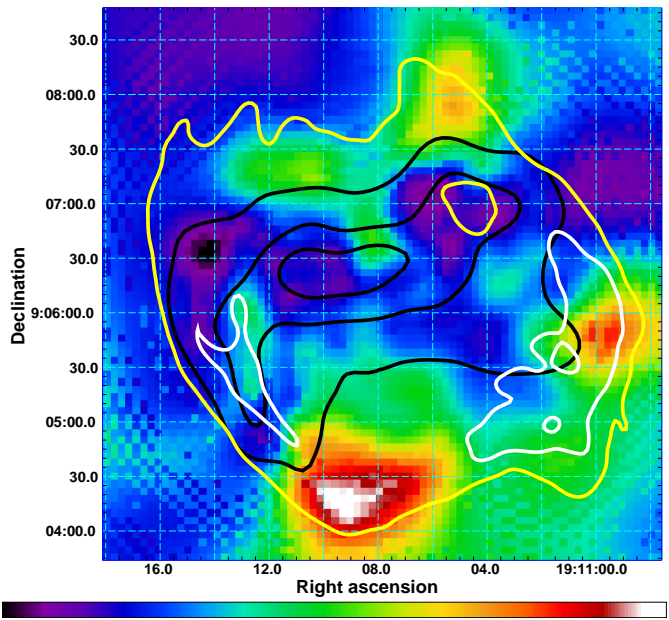,width=\columnwidth}
     }}
\caption{\emph{XMM-Newton} EPIC image of the count-rate in the $1.77-1.9$ keV energy band divided by the count-rate in the $1.96-2.06$ keV energy band. The image is vignetting-corrected, adaptively smoothed (with signal-to-noise ratio 25) and background-subtracted. The color bar ranges between 1 and 2.2. We have 
superimposed (\emph{in black}) the X-ray contour levels derived from the $1-9$ keV  EPIC image at $25\%$, $50\%$, and $75\%$ of the maximum, and the contour levels of the radio image at 327 MHz (obtained by \citealt{llk01}) at 10\% (\emph{yellow}) and 50\% (\emph{white}) of the maximum.} 
\label{fig:ratio}
\end{figure}

As shown in Paper I, although both the X-ray emitting components seem to be associated with overabundant plasma, it is not possible to rule out the possibility for the soft component to be associated with the shocked circumstellar material (CSM) or interstellar medium. The hard component is, however, associated with the ejecta.
As shown in Paper I, the Si He-like emission is clearly associated with the soft component, whose contribution becomes negligible in the energy band of the Si H-like emission line complex. 
To investigate the spatial distribution of these components and their relative weight, we produce an \emph{XMM-Newton} EPIC image of the ratio Si He-like/Si H-like emission (i. e. the count-rate image in the $1.77-1.9$ keV energy band divided by the count-rate image in the $1.96-2.06$ keV energy band). This map has been produced according to the data analysis procedure described in Paper I and is shown in Fig. \ref{fig:ratio}. The Si He-like/Si H-like ratio is enhanced in the outer parts of the remnant (with two maxima in the northern and southeastern parts of W49B), while it is almost uniform in the jet-like central region and at its eastern and western ends. If we associate the cool component (i. e. the Si He-like emission) with the ISM (or with the CSM) and the Si H-like emission with the ejecta, this may indicate that the ejecta are mainly distributed along the elongated jet-like structure, while the shocked ambient material totally embeds the remnant. On the other hand, the Si He-like/Si H-like ratio reaches its maxima in regions with low surface brightness, where the statistics are not good enough to make these features completely reliable.

At the eastern border of the remnant, where bright radio emission has been observed (see, for example, \citealt{llk01}), a shocked molecular H$_{2}$ cloud has been revealed by \citet{krr06}, who derived estimates of the temperature (few $10^{3}$ K) and density ($2-3\times10^{3}$ cm$^{-3}$) of the molecular gas. Since in Paper I we estimated for the hotter X-ray emitting component a density of a few cm$^{-3}$, the H$_{2}$ cloud is about three order of magnitude denser than the ejecta. The cloud then acts as a wall that prevents the expansion in the East direction. Molecular H$_{2}$ emission is present also in the south~western region of the shell, where we observe high Si He-like/Si H-like ratios and where copious [Fe II] and radio emission are observed.

\section{The physical origin of W49B}

\begin{figure}[htb!]
 \centerline{\hbox{     
     \psfig{figure=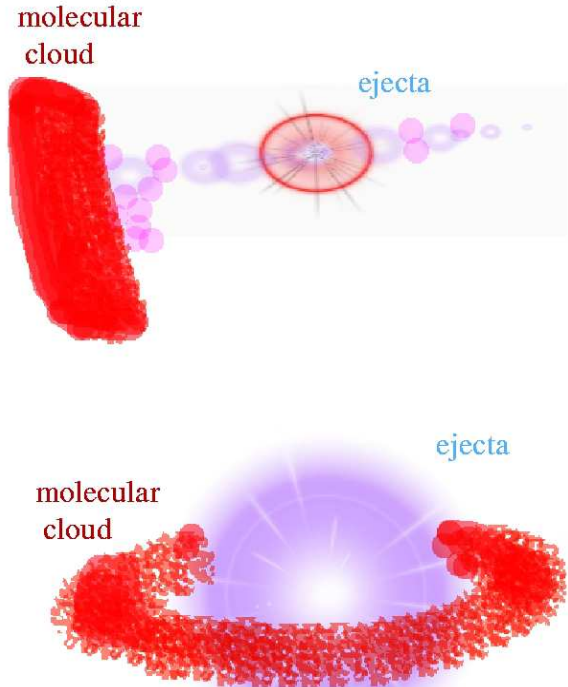,width=\columnwidth}
     }}
\caption{Schematic representation of the two possible physical scenarios associated with W49B.}
\label{fig:sketch}
\end{figure}

The peculiar jet-like morphology of W49B and the evidence for the presence of  X-ray emitting ejecta suggest that the remnant is the result of a highly aspherical bipolar explosion. On the other hand, it is also possible that W49B originated by a spherical explosion in a complex ambient medium with enhanced density in a torus-like structure all across W49B. In this case, we would observe only a small fraction of the ejecta, i. e. only the material expelled in the plane of the torus that is interacting with the reverse shock and is heated up to X-ray emitting temperature. A schematic representation of the two scenarios is shown in Fig. \ref{fig:sketch}. As we have shown in Paper I, both these interpretations are consistent with the spectral analysis results and with the observed abundance pattern. 

\begin{figure}[htb!]
 \centerline{\hbox{     
     \psfig{figure=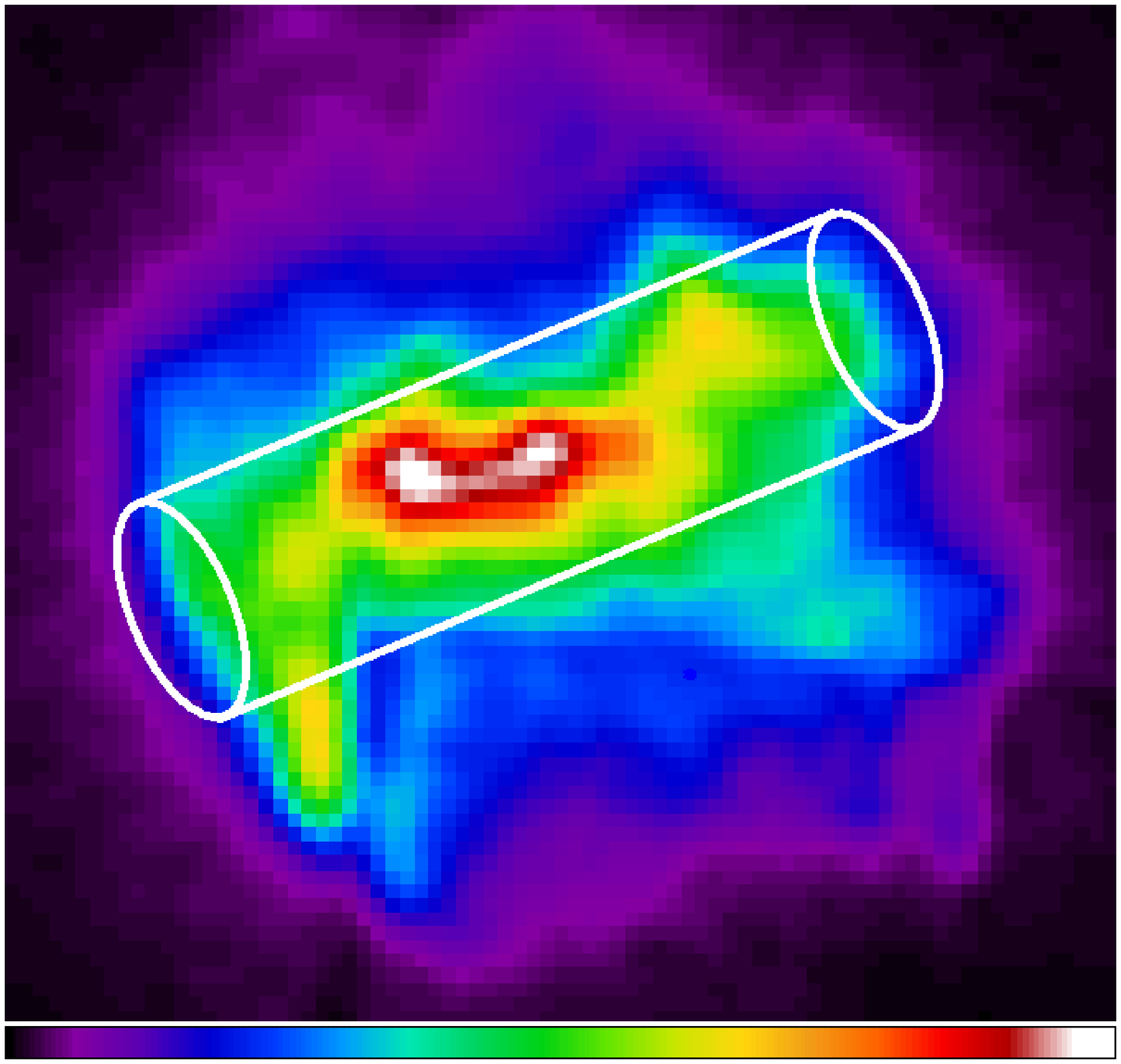,width=\columnwidth}
     }}
 \centerline{\hbox{     
     \psfig{figure=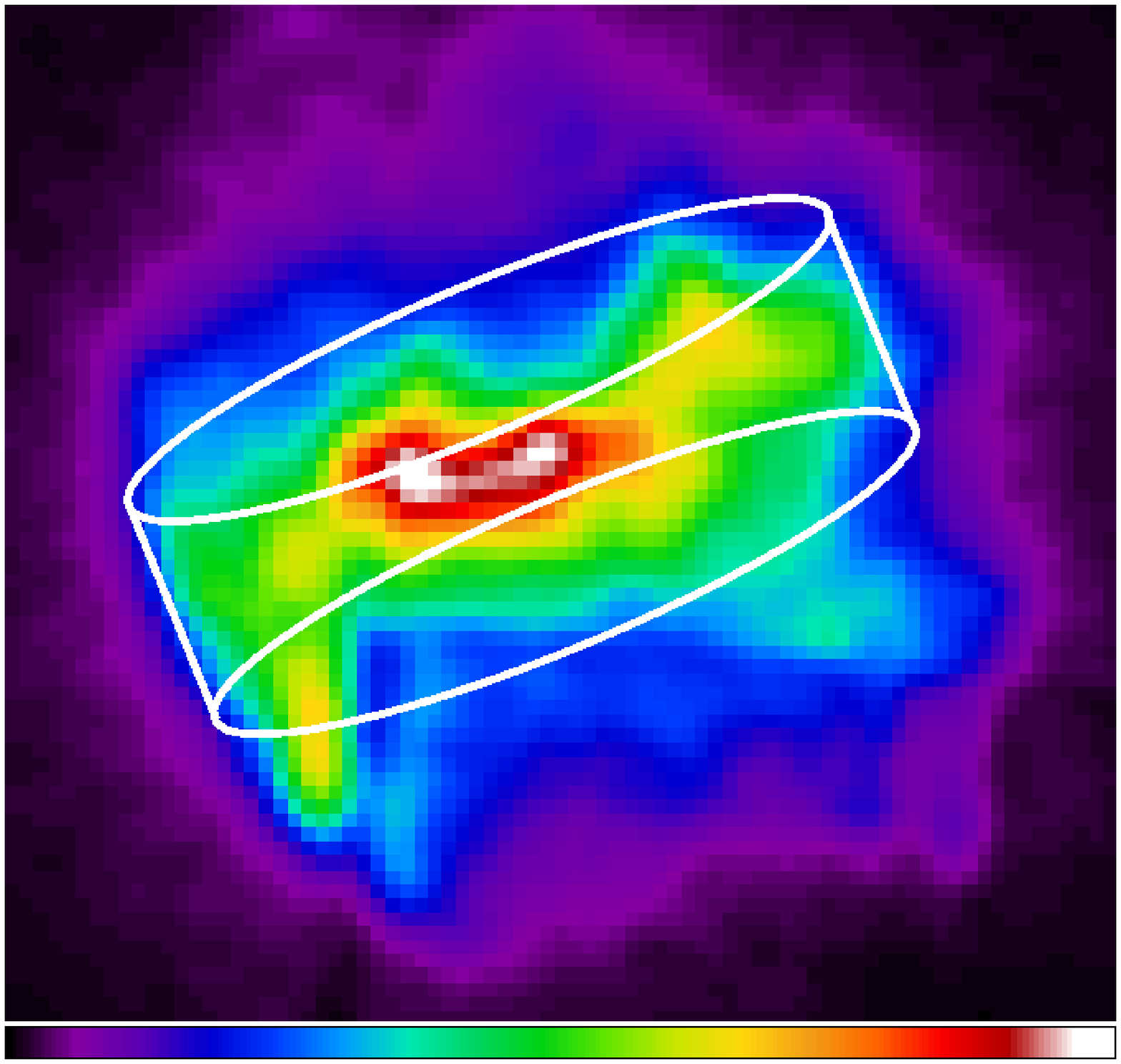,width=\columnwidth}
     }}
\caption{Same as Fig. \ref{fig:w49bspec} where we have superimposed the cylindric volumes used to estimate the mass of the X-ray emitting ejecta for a jet-like explosion (\emph{upper panel}) and for a spherical explosion (\emph{lower panel}).}
\label{fig:cilindri}
\end{figure}

We can estimate the mass of the X-ray emitting ejecta in both these scenarios. As shown in Paper I, the hard component is clearly associated with the ejecta, therefore, for the mass estimates, we will refer only to this component. Our mass estimates, however, would at maximum double by associating also the soft component with the ejecta, since the emission measure of the soft component is similar to that of the hot component (within a factor of two, see Table 3 in Paper 1). Notice also that we assume the mass density to be $\rho=\mu m_{H}n_{H}$, with $n_{e}=n_{H}$ and with $\mu=1.26$. This $\mu$ value corresponds to cosmic abundances, therefore, since in the hard component we find significant metal overabundances, our estimates can be considered as lower limits of the mass of the ejecta. In the case of aspherical jet-like explosion we can assume, for the sake of simplicity, that the ejecta are distributed into a cylinder of radius $R_{jet}=0.55'$ and height $H_{jet}=3.6'$ (shown in the upper panel of Fig. \ref{fig:cilindri}), with volume $V_{jet}\sim1.27\times 10^{57}$ cm$^3$ at 8 kpc. Since the emission measure per unit area in the central region of the remnant is $1.8\times10^{20}$ cm$^{-5}$ (cfr. Table 4 in Paper I), we derive a mean density of the ejecta $n_{jet}\sim 4.8$ cm$^{-3}$ and then a total mass $M_{jet}\sim 6$ M$_{\odot}$.
On the other hand, in the case of a spherical explosion, we would observe only the ejecta heated by the reflected shock generated by the impact of the main shock with the circumstellar torus. We can therefore assume that the ejecta are distributed into the cylinder shown in the lower panel of Fig. \ref{fig:cilindri} (with radius $R_{sph}=1.8'$, height $H_{sph}=1.1'$ and volume $V_{sph}\sim4.1\times10^{57}$ cm$^{3}$ at 8 kpc). Taking the emission measure per unit area ($EM$) of this region into account, we obtain a density $n_{sph}=\sqrt{EM/2R_{sph}}\sim2.6$ cm$^{-3}$ and an X-ray emitting mass $M_{sph}\sim 10$ M$_{\odot}$. It is also possible that the ejecta do not occupy all the cylindric volume, but only a small fraction of it. For example, we can assume the ejecta to occupy a torus with outer radius $R_{sph}$ and inner radius $gR_{sph}$ with $0<g<1$. This assumption, however, does not alter very much our mass estimate, since, in this case, we would have a mass $M_{sph}'\sim M_{sph}~(1-g^{2})/\sqrt{1-g}$, i. e., for example, $M_{sph}'=0.875 ~ M_{sph}$ for $g=0.75$. In the spherical explosion scenario, $M_{sph}$ would be only a small fraction of the total ejected mass (the fraction interacting with the reflected shock).
In fact, by assuming the ejecta to be distributed into a sphere with 1.8' radius, we obtain $M^{tot}_{sph}\sim 25$ M$_{\odot}$. This would imply a very high (maybe unrealistic) zero age main sequence mass $M_{ZAMS}$ of the progenitor star, given that, before the SN explosion, massive stars lose a very large amount of their mass ($\sim 75\%$, as shown by \citealt{dwa06}) through stellar winds. As an example, notice that the bright hypernova SN Ic 1998bw has very massive ejecta ($\sim 10$ M$_{\odot}$) and $M_{ZAMS}\sim40$ $M_{\odot}$ (\citealt{imn98}). We conclude that the aspherical explosion scenario seems to be preferable according to our mass estimates. 

Aspherical explosions are a common feature of hypernovae (i. e. supernova explosions with $E\sim10^{52}$~erg, for a review on hypernovae see \citealt{nmu03}), that are linked to $\gamma-$ray bursts, GRB, (see Paper I and references therein). In Paper I we showed that the comparison between the observed abundance pattern and the yields of the explosive nucleosynthesis models by \citet{mn03} seems to exclude the association of W49B with a hypernova in favor of a normal SN or an aspherical SN explosion with typical energy $\sim 10^{51}$ erg. Therefore, the observed abundance pattern in W49B does not favor a hypernova explosion of sufficient energy to generate a typical GRB.

An independent way to ascertain whether W49B is the result of a hypernova explosion consists in estimating the total energy of the remnant in the case of aspherical jet-like explosion. Assuming that the ejecta have been shocked by a reverse shock, we can derive the relative velocity of the ejecta with respect to the reverse shock, $v_{r}$. In fact, as shown in \citet{mh80},
\begin{equation}
T_{s}=\left( \frac{v_{r}}{839~\rm{km}~\rm{s}^{-1}}\right)^{2}\times10^7 \rm{K}~{\rm ,}
\end{equation}
where $T_{s}$ is the post-shock temperature of the plasma. The spectral analysis performed in Paper I allowed us to derive the electron post-shock temperature $T_{e}\sim2.3$ keV (see Table 4 in Paper I) which is always $\le T_{s}$. If we assume the ion temperature to be of the same order of $T_{e}$, we obtain the lower limit $v_{r}>1350$ km$/$s. We remind the reader that this is the speed of the ejecta in the reference frame of the reverse shock and is therefore larger than the speed of the ejecta in the reference frame of the unperturbed ambient ISM. In fact, given the mass of the shocked ejecta, it is reasonable to conclude that the reverse shock is moving towards the center of the remnant. Notice that this is true also if we assume the ejecta to be heated by the reflected shock generated by the interaction of the main shock front with the dense molecular cloud. The $v_{r}$ value is remarkably in agreement with the estimate of the X-ray forward shock velocity with respect to the unperturbed medium, $v_{s}\sim1200$ km$/$s, derived by \citet{krr06}. 
Taking our $M_{jet}$ estimate into account, we then derive a kinetic energy $K_{ej}\sim1.2 \times 10^{50}$ erg. 
Notice that, according to \citet{rak05}, it is reasonable to assume that $0.1<T_{e}/T_{s}<1$ (for shock velocity $<5000$ km/s). In the case $T_{e}=0.1~T_{s}$, we then obtain the upper limit $v_{r}< 4300$ km/s and then $K_{ej}<1.2 \times 10^{51}$ erg.
The internal energy $U_{jet}=(3/2) n_{jet}kT_{s}V_{jet}$ is in the range $3.4\times 10^{49} - 3.4\times10^{50}$ erg (for $T_{e}/T_{s}=0.1-1$) and is even lower. The total energy of the remnant\footnote{In the case of the spherical explosion scenario we find $K_{sph}\sim2\times 10^{50}- 2\times10^{51}$ erg and $U_{sph}\sim 6\times 10^{49}- 6\times10^{50}$ erg.} is then, in any case, lower than $1.5\times 10^{51}$ erg. Although our estimates of mass and velocity are quite rough, it would be necessary for the velocity of the ejecta to be one order of magnitude larger than $v_{jet}$, or for the total mass to be two orders of magnitude larger than $M_{jet}$ to obtain $10^{52}$ erg. This result, together with the observed abundance pattern, concur in excluding the possible association of W49B with a hypernova (and a GRB), although this association cannot be definitely ruled out especially given the emergence of the class of underenergetic GRBs associated with SNe (\citealt{mdn06}).


\section{Conclusion}
\label{Conclusion}

The morphology and physical properties of the X-ray emitting plasma in W49B have been extensively studied in Paper I, where we have shown that two thermal components in collisional ionization equilibrium provide a good description of the \emph{XMM-Newton} spectra.
While the results of the spectral analysis do not allow us to clearly distinguish between solar and enhanced abundances for the soft component, the map of the Si He-like/ Si H-like emission lines presented here suggests an identification of the soft component with CSM material.

We have also shown indications that W49B is linked to an aspherical supernova explosion in a wind cavity, bounded on the eastern side of the remnant by the infrared emitting molecular clouds. It is also possible that W49B is the result of a spherical SN in a pre-existing structure in the ambient environment, maybe generated by the strong stellar winds of the progenitor star, which may have produced an aspherical reverse shock. 
However, our estimates of the ejected mass in the case of a spherical explosion yield to very high values, while more realistic values can be obtained assuming a jet-like morphology of the ejecta. 

There is a link between very energetic jet-like explosions (or hypernovae) and $\gamma-$ray bursts, although a bipolar explosion does not necessarily imply a GRB, because a high explosion energy (some $10^{52}$ erg) is also required.
The abundance analysis presented in Paper I has shown that the association of W49B with a GRB is not supported by observational evidence and the estimates of the total energy of the ejecta presented here confirm this conclusion.

W49B seems then to be the result of an aspherical jet-like supernova explosion explosion of normal energy. Notice that, in this case, we would have significant differences between the eastern arm of the jet, where high values of temperature and abundance have been found, and the western arm. It is, however, difficult to ascertain whether this East-West anisotropy is the result of an anisotropic explosion (like, for example, that produced by the instability of the accretion shocks described by \citealt{bm06} and \citealt{bmd03}), or by the interaction of the jets with a complex, inhomogeneous ambient medium. 

\section*{Acknowledgements}
We thank the referees and J. Vink for useful suggestions. This work was partially supported by the Ministero dell'Istruzione, dell'Universit\`{a} e della Ricerca and by ASI-INAF contract I/023/05/0.








\end{document}